\documentclass[12pt]{article}

\usepackage{amsmath,amsthm,amssymb}
\usepackage{boxedminipage}

\usepackage{graphicx}

\usepackage{pgfplots}
\usepackage{remark}
\usepackage{float}
\usepackage{pgfplotstable}
\pgfplotsset{compat=newest}

\begin{document}

\title{\bf A context-aware e-bike system to reduce pollution inhalation while cycling }
\author{Shaun Sweeney, Rodrigo Ordonez-Hurtado, Francesco Pilla,\\ Giovanni Russo, David Timoney, Robert Shorten\protect\footnote{Rodrigo Ordonez-Hurtado, Francesco Pilla, David Timoney and Robert Shorten are with University College Dublin, Shaun Sweeney and Giovanni Russo are with IBM Research. Contact robert.shorten@ucd.ie.}\protect\footnote{The results documented in this paper were presented at the 18th Yale Workshop on Adaptive and Learning Systems (https://www.eng.yale.edu/css/). A printed copy of a preliminary version of this paper was presented, without copyright transfer, to the workshop participants. A version of this paper was submitted to IEEE Transactions on Intelligent Transportation Systems on 4th August 2017.}}

\date{}

\maketitle

\begin{abstract}
The effect of transport-related pollution on human health is fast becoming recognized as a 
major issue in cities worldwide. Cyclists, in particular, face great risks, as they typically are most exposed 
to tail-pipe emissions.  Three avenues are being explored worldwide in the fight against urban pollution:
(i) outright bans on polluting vehicles and embracing zero tailpipe emission vehicles;
(ii) measuring air-quality as a means to better informing citizens of zones of higher pollution; and
(iii) developing smart mobility devices that seek to minimize the effect of polluting devices on citizens as they transport goods and individuals in our cities.  
Following this latter direction, in this paper we present a new way to protect cyclists from the effect of urban pollution. Namely, by exploiting 
the actuation possibilities afforded by pedelecs or e-bikes (electric bikes), we design a cyber-physical system that 
mitigates the effect of urban pollution by indirectly controlling the minute ventilation (volume of air inhaled per minute) of cyclists in polluted areas. 
Results from a real device are presented to illustrate the efficacy of our system.
\end{abstract}

\section{Introductory remarks}

Our objective in this work is to exploit the actuation possibilities afforded by e-bikes to deliver new services to cyclists. The specific focus is to use these actuation possibilities to indirectly control the breathing rate of the cyclist with a view to reducing their intake of harmful tail-pipe emissions. We do this by providing judiciously, electrical assistance to the cyclist in areas where pollution is highest.\newline

Recent work in the area of addressing transport-related pollution has focused in three main areas: (i) building vehicles that do not pollute, such as electric vehicles (EVs) \cite{Shorten2017}; (ii) low-cost ubiquitous urban sensing of pollution with a view to informing people of dangerous pollution present in their surroundings; and (iii) using smart devices that adapt to their surroundings to protect humans, such as hybrid actuation in plug-in Hybrid Electric Vehicles (PHEVs). Our present work is most related to this latter thread of research. For example, PHEVs have two modes of operation: a fully electric mode, and a hybrid mode, the latter of which is designed by the manufacturer to maximize fuel efficiency \cite{stockar2011energy, barsali2004control}. Recently, several authors such as \cite{schlote2013cooperative, IJCSPONGE, naoum2016smart} have suggested exploring the actuation possibilities in such vehicles, namely to automate the on/off switching of the fully electric mode, to address not only fuel efficiency but also pollution issues in urban areas. These ideas are further explored for a network of vehicles in \cite{hausler2014framework} by formulating a constrained optimization problem. More recently, these ideas have been extended with a particular focus on pedestrians and cyclists \cite{annie, adam}.  Finally, it is worth noting that our work is also related to conventional management strategies for PHEVs; see \cite{stockar2011energy}. \newline

Specifically, in this work, our aim is to  indirectly control the total volume of air inhaled per minute (hereto referred to as the {\em ventilation rate}) of a cyclist by judiciously applying electric motor assistance.  To do this, we instrument and modify an off-the-shelf electric bike, and design a cyber-physical control system to manage the interaction of the cyclist and electric motor. We apply our system to mitigate the effects of urban pollution on a cyclist. Finally, we perform a number of tests on real subjects to demonstrate the feasibility of the system.
A video demonstrating our proposed system can be found at \cite{VideoDemo}. 

\section{Human health}\label{health}

The link between particulate matter (PM) and human health has been the subject of a number of recent studies. PM is a generic term used for a type of pollutants that consist of a complex and varied mix of particles suspended in air. The size of the PM particles varies, with PM$_{x}$ defining a category of PM with aerodynamic diameter smaller than $x$ $\mu$m. Other definitions for these pollutants include referring to them as ultra-fine, fine or coarse particles, which is again based on particle size. The major components of PM are metals, organic compounds, material of biological origin, ions, reactive gases and the particle carbon core. There is strong evidence to suggest that ultra-fine and fine particles are more harmful to human health than larger ones as these particles can travel further into the respiratory tract. Coarse particles deposit mainly in the upper respiratory tract but fine and ultra-fine particles can travel further and reach the lung alveoli \cite{Kampa2008}. Seaton's 1995 paper \cite{Seaton1995} published in the Lancet is in agreement with the harmful effects of PM noting that {\it``epidemiological studies have consistently shown an association between particulate air pollution and not only exacerbations of illness in people with respiratory disease but also rises in the number of deaths from cardiovascular and respiratory disease among older people'}'. It is hypothesized that ultra-fine particle characteristics of air pollution provoke alveolar inflammation, which causes changes in blood coagulability (ability of the blood to clot) and release of mediators able to provoke attacks of acute respiratory illness. These blood changes result in an increase in the exposed population's susceptibility to acute episodes of cardiovascular disease. Of particular note, a recent study by Chen \cite{Chen2017} also published in the Lancet found that living near major roads was associated with a higher incidence of dementia. For further work in this direction focussed on pedestrians - see \cite{Sin2017} and for further reading, see  \cite{ Brunekreef2002}. 

\subsection{Cycling and air pollution} \label{sec:lr_cycling_pollution}	
Panis's 2010 paper \cite{IntPanis2010} investigated the relationship between the amount of PM that cyclists inhaled compared to car passengers. The study was conducted in three different Belgian regions--Brussels, Louvain-la-Neuve and Mol. Subjects in the study were first driven by car and then cycled along identical routes in a pairwise design with the aim of comparing lung deposition of particle number concentrations (PNC) and PM between car trips and biking trips. Atmospheric concentrations of PNC and PM measurements were similar for both cycling and car journeys across all three locations. However, breathing frequency, breathing volume and journey time were all greater for cyclists than for cars. The study also made use of the result from \cite{Daigle2003} which showed that lung deposition fractions increases strongly with exercise. This meant that even though the atmospheric concentrations were similar, quantities of particles inhaled by cyclists were between 400-900\% higher compared to car passengers on the same route. The longer duration of the cycling trip also increased the inhaled doses. Recent Bigazzi's studies \cite{bigazzi_breath_2016,AlexanderYBigazzi2017} aimed to determine (i) biomarkers that could be used to assess a cyclist's uptake of pollutants in urban environment, and (ii) optimal cycling speeds to minimize a cyclist's inhalation of those biomarkers. The work in \cite{bigazzi_breath_2016} aimed to determine a way to assess the inhalation of harmful pollutants by cyclists for trips through different kinds of areas. The study identified 26 volatile organic compounds (VOCs) such as CFCs, benzene, styrene and carbon disulfide, and compared the amounts of these compounds that were present in ambient air with the amounts that were present in a cyclist's breath after cycling through a given area. The study identified 8 of these VOCs as being potentially useful breath biomarkers, and found statistically significant increases (compared to background levels) in the concentration of these biomarkers in the breath of cyclists after cycling in high-traffic streets and industrial areas.

\section{The e-bike and cyber-physics}\label{smartbike}

We now discuss the design of our e-bike based-pollution mitigation system. An electric bike is very similar to a standard bike, with one small difference--it also has an electric motor that can assist the cyclist in completing journeys. From a control perspective, the electric motor can be used to reject disturbances (wind, hills), and to provide new services to the cyclist.
Our basic objective here is to develop a context aware e-bike system that can provide a level of electrical assistance in response to context. In particular, our specific focus is to enable strategies to mitigate the effect of pollution on cyclists (and also to address an energy management issue related to topology and wind disturbance rejection). The system development involves three main steps: (i) bespoke modification of a standard e-bike to enable context aware behavior; (ii) the development of low-level control and optimization algorithms to be deployed on the e-bike; and (iii) the integration of these algorithms with an Android-based route prediction engine.\newline

{\bf Comment:} We remark here that the system we propose in this paper cannot be fully studied and designed if the interfaces (or interactions) between the human and the algorithmic procedures running on the e-bike are not fully understood or neglected when the system is modeled. Systems arising from the integration and interconnection of continuous processes and procedural/algorithmic processes are known as cyber-physical systems (CPSs) \cite{Lee_08,Lee_15}.

\begin{figure}[h]
	\begin{center}
		{\includegraphics[width=5in]{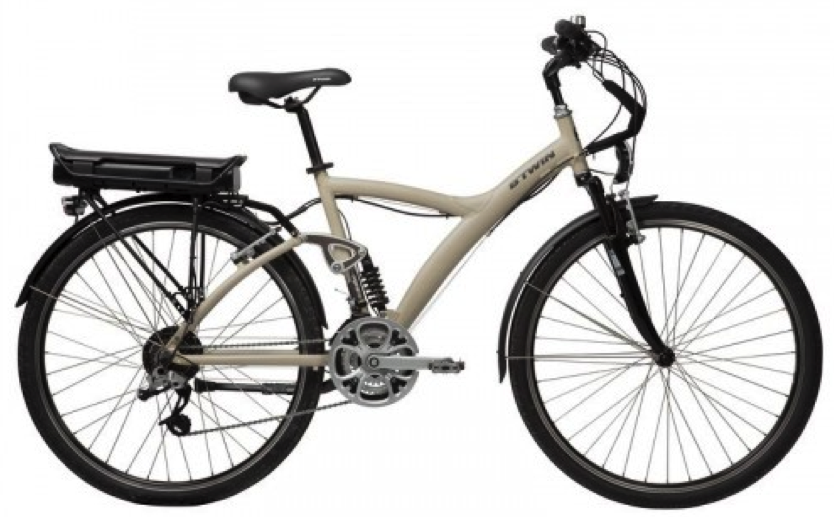}}
		\caption{Electric bike from BTwin.}
		\label{ebike}
	\end{center}
\end{figure}

\subsection{Hardware design}

The e-bike that we use is a modified 
BTwin Original 700 purchased from Decathlon (see Figure \ref{ebike}). The original bike is 
equipped with a Samsung Li-ion 36V battery and a Bafang controller, but it was modified in our study in several ways.
First, to facilitate control design, the original motor controller was replaced by a more advanced controller: a Grinfineon C4820-GR.
Second, several measurement sensors were added to the bike; these include sensors to measure: pedal torque and speed (using a THUN X-CELL RT sensor\footnote{http://www.ebikes.ca/shop/electric-bicycle-parts/torque-sensors/thun-120l.html}), battery voltage, motor current,
wheel speed, motor temperature,  brake sensor, and hand throttle sensor.
These are used to derive information pertaining to state-of-charge of the battery, electrical power input to motor, distance traveled, human power input, and acceleration.
Data from sensors are read either by using an Arduino (brake and hand throttle sensors) or using a commercially available computer system--the Cycle Analyst\protect\footnote{www.ebikes.ca}--(all the other sensors), and then communicated to a smartphone using a 
bespoke especially designed Arduino-controlled Bluetooth module. The main reason for the data transfer to the smartphone is to exploit
the convenience of remotely controlling the motor, and to exploit external data streams through the Internet connection
on the smartphone. Control inputs are finally sent to the bike controller using the same Bluetooth based communication system. Full details of the hardware system are given 
in \cite{Shaun2017}.

\subsection{Algorithmic components}

\begin{figure}[h]
	\begin{center}
		{\includegraphics[width=5in]{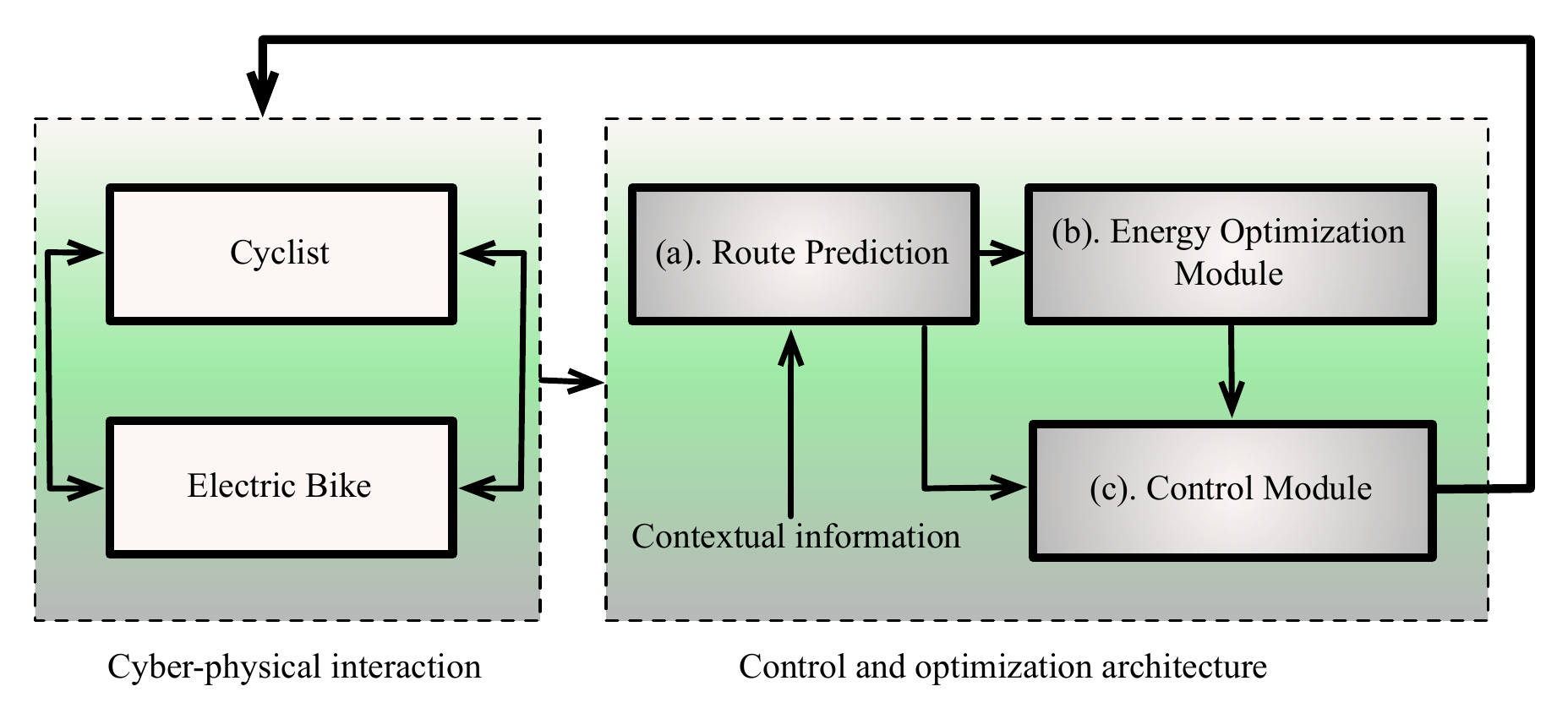}}
		\caption{Algorithmic architecture of e-bike system.}
		\label{fig:architecture}
	\end{center}
\end{figure}

We now outline the main algorithmic components of the e-bike system. The algorithmic architecture is depicted in Figure \ref{fig:architecture}.

Historical data from the cyclist is used to model both the intent of the cyclist and likely energy consumption along predicted routes. This module 
is also used to parse information available for decision support. Viewing the battery 
state-of-charge as a limited resource, an energy management module is then used to optimally calculate the   
allowed energy consumption along the expected route to maximize some utility and/or to recommend routes. Finally, a control module manages the interaction between 
the cyclist and the electric motor. We now briefly give an overview of each of these modules.

\subsubsection{Route parsing engine}

The goal of this module is that of predicting the likely route that the cyclist is going to take. The algorithm presented in this Section is a basic version of a set of route prediction algorithms that recently appeared in the context of connected vehicles in \cite{Sam,annie}. Essentially, this module takes as input the past trips of the cyclist and, based on his/her past trips, it outputs the most likely route that the cyclist is going to take. In this context, a trip is defined as a sequence of road segments. We denote the $k$-th road segment by $r_k$. Assume now that the cyclist is traveling along the road segment $r_j$. Given the past history of the cyclist, the road segment $r_j$ will belong to a number of different routes, say $\mathcal{R}_1^j,\ldots, \mathcal{R}_k^j$. We also denote by $N_t^j$ the total number of times that route $\mathcal{R}_t^j$ was taken in the past, when the bike was on segment $r_j$.  Based on these data, a probability, say $p(\mathcal{R}_t^j)$ is computed for each of these routes as follows:
\begin{equation}
p(\mathcal{R}_t^j) = \frac{N_t^j}{\sum_{i=1}^k N_i^j}, \forall t \in \left\lbrace 1,2,\ldots, k \right\rbrace.
\end{equation}
Then, the route predicted by the algorithm is the one that corresponds to the highest $p(\mathcal{R}_t^j)$. We also note that the probability $p_s^i$ of taking segment $r_s$ in the future can also be simply computed as the sum of the probabilities of all routes that contain that road segment. The probability $p_s^i$ is given as an input to the optimization module described in the next Section.

\subsection{Energy budget management}

We now formalize the optimization problem:
\begin{equation}\label{optproblem}
\mbox{Optimization Problem:  }\left\{
\begin {array}{l}
\underset{x^i_s}{\max} \quad \sum_{s \in S} p^i_s d^i_s \overline{e}_{s} x^{i}_{s}\\
{\text{s.t.}} ~ \sum_{s \in S} p^i_s \overline{e}_{s} x^{i}_{s}  \leq E^i_{av}\\
\quad 0 \leq x^{i}_{s} \leq 1
\end{array},
\right.
\end{equation}
which has the goal of managing in an optimal way the usage of the available electric battery energy (as in \cite{annie}). We denote by:
\textit{(i)} $\mathcal{S}$ the set of all road segments from the historical routes taken in the past by the cyclist;
\textit{(ii)} $\overline{e}_{s}$ the expected battery consumption along the $s$'th segment of $\mathcal{S}$;
\textit{(iii)} $d_s^i$ the expected pollution level along segment $s$;
\textit{(iv)} $E^i_{\textrm{av}}$ an available energy budget when the bike is located at the $i$-th segment;
\textit{(v)}  $x_s^i \in \left[0, \ 1\right]$ the set of decision variables that we wish to optimally compute.
In particular, $x_s^i$ is the percentage of power that the e-bike provides to the cyclist on each segment $r_s \in \mathcal{S}$. The superscript $i$ indicates that an optimal prediction is performed when the {bike is traveling} on the road segment $r_i$, and a new prediction will be performed when the {bike} enters a new road segment, when a new (possibly more accurate) prediction of the route will be performed. Roughly speaking, the Optimization Problem (\ref{optproblem}) aims at ensuring electrical energy is available in areas corresponding to high pollution levels. This optimization incorporates uncertainty of the route, by giving more importance to the most likely routes. We also remark that the optimization problem is solved iteratively every time the bike enters a new road segment.

\subsection{Control}

\subsubsection * {A. Basic modeling}

\begin{figure}[h]
	\begin{center}
		{\includegraphics[width=5in]{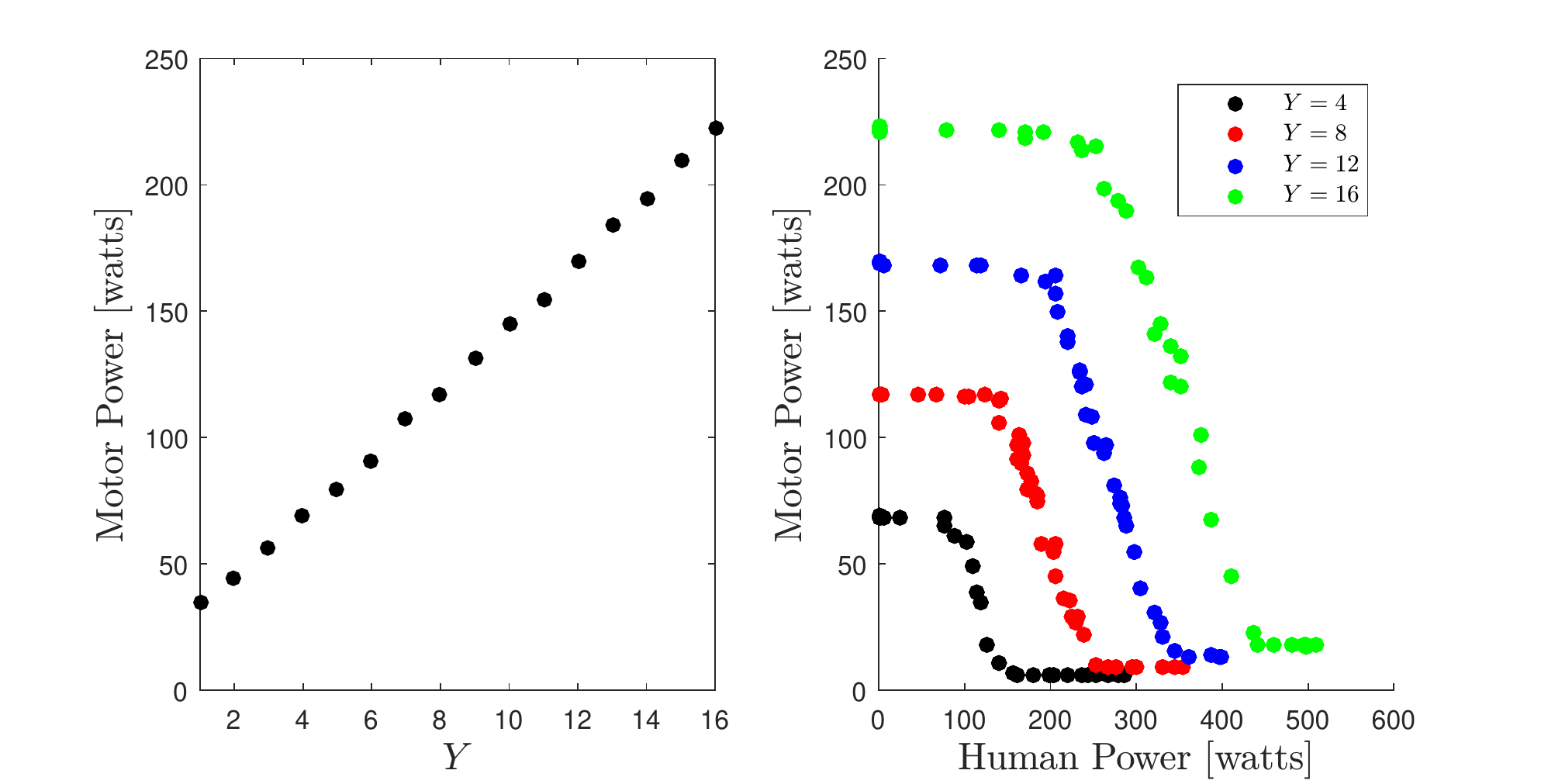}}
		\caption{Impact of $Y$ on the input power of the motor for zero (left) and non-zero (right) human power.}
		\label{fig:MotorPower}
	\end{center}
\end{figure}

In what follows, we ignore any dynamics associated with the e-bike's electric motor (hereafter just referred as the {\it motor}), 
and analyze the system based on measurements with fixed gear configuration. Let $Y$ denote the control input to the motor, where $I_{M}=\mu Y$ is the motor current and $\mu$ is a proportionality constant. To evaluate the impact of $Y$ on the motor in the cases of having zero and non-zero human power input, 
some definitions must be introduced. First, the instantaneous input power to the motor, $P_{M_{in}}$, is
defined as
\[
P_{M_{in}}\left[watt\right]=V_{M}\left[volt\right]\times I_{M}\left[amp\right],
\]
where $V_{M}$ is the motor voltage. Second, the instantaneous input power from the human to the bike, $P_{H_{in}}$, is defined as
\[
P_{H_{in}}\left[watt\right]=\tau_{p}\left[newton \; meters\right]\times\omega_{p}\left[rad/sec \right],
\]
where $\tau_{p}$ is the torque provided by the cyclist at pedals, and $\omega_{p}$ is the angular velocity at pedals. Note that, in the case of zero human-power input, the motor power will only depend on the value of the control input $Y$ (as $I_M$ is proportional to the demanded torque). However, in the case of non-zero human-power input, the torque provided by the motor will depend on the load seen by the motor, which clearly changes as a function of the power provided by the cyclist; thus, in this case, the motor power will depend on both $Y$ and $P_{H_{in}}$. This is illustrated in Figure \ref{fig:MotorPower}.

The power $P_{w}$ required to move the rear wheel can
be approximated as
\[
P_{w}\approx P_{M_{out}}+P_{H_{out}},
\]
namely, as the sum of the output motor power
\[
P_{M_{out}}=E_{m}P_{M_{in}},
\]
where $E_{m}$ is the efficiency of the motor, and the output human
power
\[
P_{H_{out}}=E_{c}P_{H_{in}},
\]
where $E_{c}$ is the efficiency of the crankset. For practical purposes, we will use the approximation
\[
P_{w}=\tilde{P}_{M_{out}}+\tilde{P}_{H_{out}},
\]
where $\tilde{P}_{M_{out}}$ and $\tilde{P}_{H_{out}}$ are the corresponding
filtered versions of $P_{M_{out}}$ and $P_{H_{out}}$  to account for 
biases and uncertainties in the system operation. Details for the aforementioned filters are provided in \cite{Shaun2017}.

\subsubsection*{B. Low-level tracking control}\label{S:Low-level-controller}

A variable of interest for control is the
proportion $m$ of power provided by the cyclist to move the rear wheel,
this is
\[
m=\frac{\tilde{P}_{H_{out}}}{P_{w}}.
\]
Thus, $m=0$ only when the system
is in full electric mode, and $m=1$ only at full human mode. 
Here, we propose a simple control approach to account for regulating
$m$ around a given reference value $m^{*}$. This is important
as, in a variety of situations such as traveling across a highly
polluted area, the variable $m$ must be maintained as low as
possible. Note that the interaction between the cyclist and the 
motor must be managed carefully. For example, $m = 0$ is illegal 
in some countries. Further, the interaction between the cyclist and motor
may be competitive rather than collaborative (the natural tendency 
for a cyclist is to increase effort in the presence of assistance). 
For control purposes, an average value of $m$
is preferred for regulation purposes due to the stochasticity 
of the cyclist. 
Thus, $m$ is proposed to be calculated as
\[
m=\frac{\bar{\tilde{P}}_{H_{out}}}{\bar{\tilde{P}}_{M_{out}}+\bar{\tilde{P}}_{H_{out}}},
\]
where $\bar{\tilde{P}}_{H_{out}}$ and $\bar{\tilde{P}}_{M_{out}}$
are the corresponding moving averages of $\tilde{P}_{H_{out}}$ and
$\tilde{P}_{M_{out}}$. Finally, a simple controller is proposed
to regulate $e_{k}=m^{*}-m$ through the updating
equation
\begin{equation}\label{eq:UpdatingRule}
Y_{k+1}=Y_{k}-\gamma e_{k},
\end{equation}
where $\gamma>0$ is a proportional gain. 

\subsubsection*{C. High-level control - setpoint generation}

The basic idea here is to manage the interaction between the motor and the cyclist to emulate a natural interaction. A number of strategies are attractive to manage
this interaction. For the purpose of this paper, we implement a filter to emulate a consensus strategy. This is given by
\begin{equation}\label{eqn:consensus}
m^{*}\left(k+1\right) = \alpha m^{*}\left(k\right) + \left(1-\alpha\right)m^{*}_{M}\left(k\right),
\end{equation}
where $m^{*}_{M}:= 1 -m^\ast $ is the setpoint for the proportion of the power provided by the motor and $\alpha$ is a control gain that can be tuned to determine the convergence rate of the algorithm. Thus, from (\ref{eqn:consensus}) we get:
\[
m^{*}\left(k+1\right) = \left(1-\alpha\right) + \left(2\alpha-1\right)m^{*}\left(k\right).
\]
The above discrete-time dynamics is stable if the control gain, $\alpha$, is positive and smaller than $1$, i.e. $0< \alpha <1$. In particular, in this case, only one equilibrium point for the above dynamics exists and this is equal to $0.5$. This means that, independently on the initial conditions, the time evolution of $m^\ast(k)$, governed by (\ref{eqn:consensus}), converges to the value $0.5$. By definition of $m_M^\ast$, this in turn implies that the dynamics (\ref{eqn:consensus}) drives the system towards the agreement value $m^{*}\left(k\right) = m^{*}_{M}\left(k\right) = 0.5$. Note that (\ref{eqn:consensus}) can be easily modified to allow tracking of a reference set point, say $o\left(k\right)$, as follows:
\[
m^{*}\left(k+1\right) = \left(2\alpha-1\right)m^{*}\left(k\right) + 2\left(1-\alpha\right)o\left(k\right).
\]
The above dynamics has equilibrium points at the constant values of a piecewise constant signal $o\left(k\right)$. We also remark here that the reference signal $o(k)$ can be provided as an output from the optimization strategy described above.\newline

\begin{figure}[h]
	\begin{center}
		{\includegraphics[clip, trim=0cm 0.4cm 0cm 0cm, width=5in]{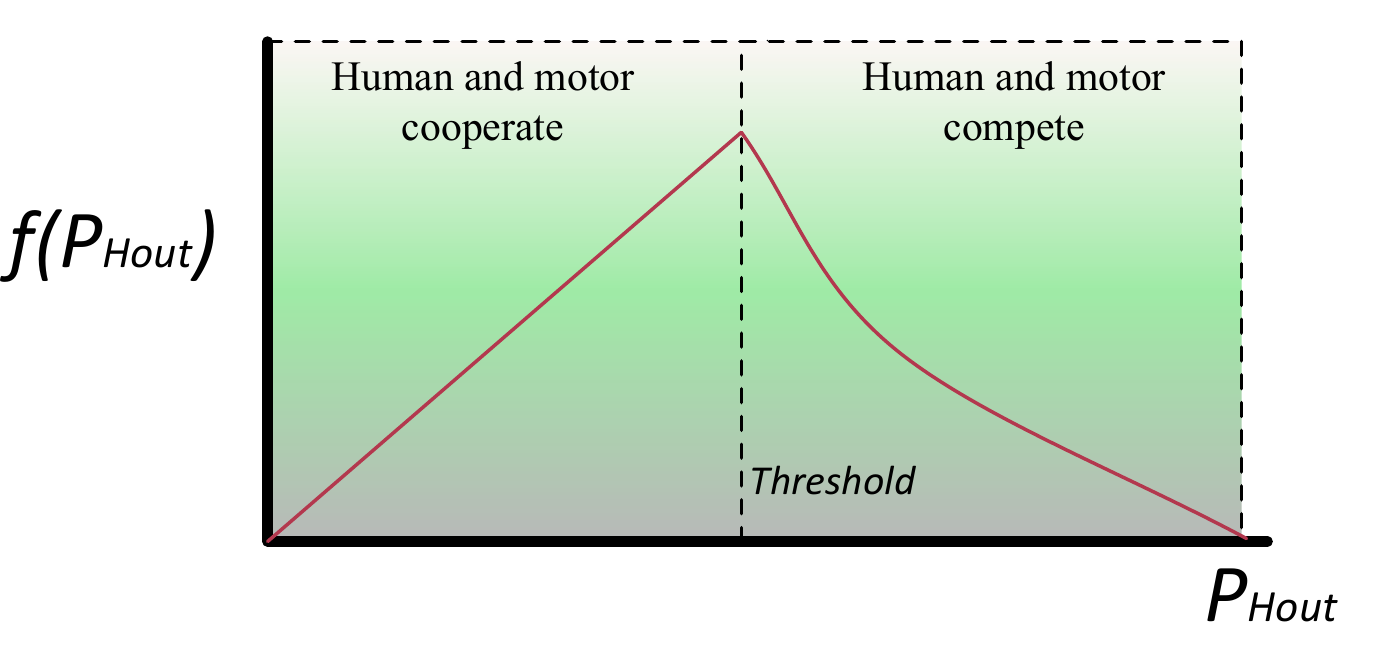}}
		\caption{The logic to specify cooperative and competitive behavior of the controller.}
		\label{fig:coop}
	\end{center}
\end{figure}

{\bf Comment:} This set-point strategy is clearly quite basic and only accounts for the interaction between the cyclist and 
the motor in a rudimentary manner. A design that takes into account this interaction in a principled manner is more difficult, mainly
due to the fact that the natural inclination of the cyclist is to increase effort as the electric motor assumes some of the torque generation responsibilities. 
In such circumstances, one wishes the motor to switch-off, and only become active when the cyclist is cooperating with the motor. Thus, any design must 
account for situations where the cyclist cooperates with the motor, and other situations when the cyclist competes with the motor. 
Motivated by the ideas in \cite{narendra} in which unstable (competing) systems stabilize each other, 
we are also exploring an alternative strategy given by
\begin{eqnarray}\label{eq:bifurcation}
\dot{P}_{{M}_{out}} = f(P_{{H}_{out}})P_{{M}_{out}} - P_{{M}_{out}}^3,
\end{eqnarray}
which is the normal form of a pitchfork bifurcation \cite{Ku:98} and where the function $f(\cdot)$ is designed so that $f(0) = 0$ and $f(\chi) > 0$, $\forall \chi >0$. A typical $f(\cdot)$
if depicted in Figure \ref{fig:coop}. The idea of using the dynamics (\ref{eq:bifurcation}) to regulate the interaction between the cyclist and the controller comes from the fact that if $P_{{H}_{out}} = 0$, then $P_{{M}_{out}}$ converges to $0$, independently on the initial conditions. On the other hand, if $P_{{H}_{out}}>0$, a bifurcation is experienced and hence $P_{{M}_{out}}$ converges to $\sqrt{f(P_{{H}_{out}})}>0$, independently on the initial conditions.
Note that in (\ref{eq:bifurcation}) the dynamics of $P_{M_{out}}$ depend on $P_{H_{out}}$, and that the dynamics of ${P}_{H_{out}}$ directly depend on the behavior of the cyclist, and the manner in which the cyclist interacts with the electric motor. Here, the assumption is that when the cyclist {\em competes} with the motor then $P_{H_{out}}$ is above some value that depends on an external variable such as route choice, pollution levels, etc.  Thus, $f(\cdot)$ is chosen so that the motor assists when the cyclist {\em cooperates} with the control strategy, and switches off in a gradual manner otherwise. 

\section{Results and measurements}\label{results}

A number of preliminary tests were conducted to empirically validate the hypothesis underlying of our 
pollution mitigation system. These are preliminary tests, based on two test subjects with a good level of fitness; no 
claim is made with regard to their statistical significance. 
The tests are merely provided in support of our {\em proof-of-concept} system.\newline 

{\bf Nomenclature:} Before proceeding, we state the meaning of the following terms that we use in the 
sequel:
(i) {\em heart rate} is the number of heart beats per minute;
(ii) {\em ventilation rate} is the total volume of air inhaled by a subject  and is measured in liters per minute;
(iii) {\em metabolic steady state}, for a constant cycling speed and constant level of electrical assistance, denotes time periods where heart rate and 
ventilation rate are approximately constant;
(iv) {\em cyclist response time} is the time needed for a cyclist to reach steady state following a step change in $m^*$.\newline

{\bf Comment:}  Note in what follows, we are verifying, empirically, the efficacy of the control system with respect to the indirect control of ventilation rate. We are not verifying the other aspects of the algorithmic system architecture (route prediction and optimization). This latter validation, as well as large population studies, will be presented elsewhere. For the study presented in the sequel, test data from the prototype system was obtained from two of the authors of the paper, who give full permission for publication of this data.

\subsection{Measurement equipment}

Ventilation rate measurements were conducted using a COSMED Spiropalm 6MWT. This device provides the ability
to measure ventilation patterns during cycling  with a fully integrated pulse oximeter to monitor blood oxygen saturation (SpO2)
and heart rate during the test. The device consists of a turbine flowmeter connected to a silicone face mask with
head cap for measurement of ventilatory parameters. The device is portable and meets ATS/ERS standards
for spirometry (2005) and 6MWT (2002) testing.
The performance of the system (bike and cyclist) was assessed in both (i) an indoor lab environment where the bicycle was mounted in a commercially available test-stand in order to simulate outdoor route profiles, and (ii) on a real road in an outdoor environment.\newline 

\subsection{Controller verification}

\begin{figure}[h]
	\begin{center}
		{\includegraphics[width=4in]{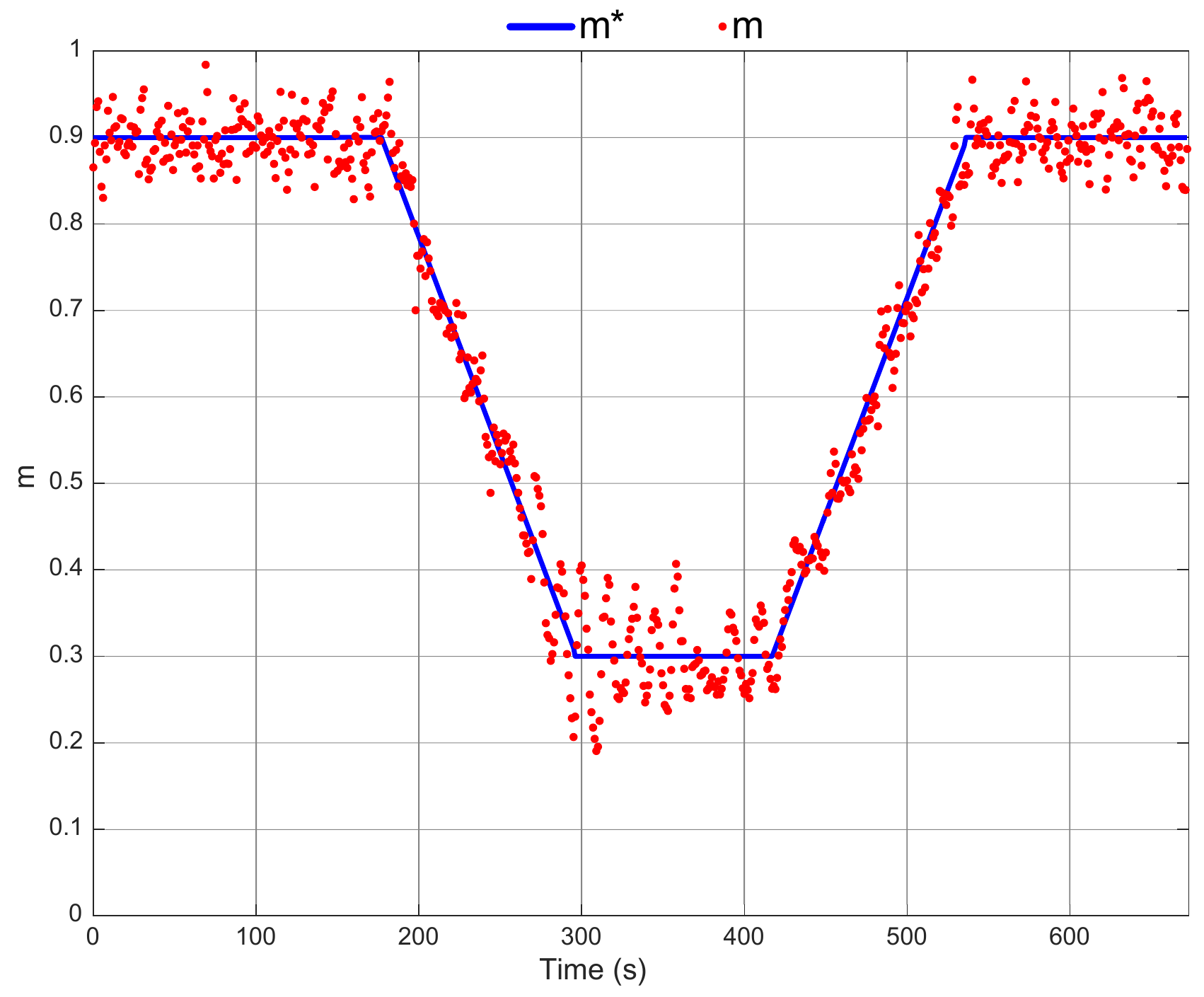}}
		\caption{Controller response.}
		\label{fig:spc}
	\end{center}
\end{figure}

A number of controlled experiments were conducted to verify the operation of the low-level control system. 
Figure \ref{fig:spc} depicts an experiment in which the cyclist was instructed to cycle at a constant 20 km/h using the
indoor testing system. The reference value of $m^*$ was then varied between $0.9$ and $0.3$ periodically over a period of 700 seconds.
As can be seen from Figure \ref{fig:spc}, the motor controller clearly regulates $m$ around the reference value.\newline 

{\bf Comment:} There is some variability in the reference tracking due to the inability of the cyclist to precisely 
maintain a constant speed. These deviations, and their interaction with the integral control system, manifest themselves 
as oscillations in the tracking signal. Note that perfect tracking is not required due to the inherent stochasticity in the cyclist's behavior.

\subsection{Ventilation rate and $m^{*}$}

\begin{figure}[h!]
	\begin{center}
		{\includegraphics[clip, trim=2.2cm 2.2cm 2.2cm 2.2cm, width=5in]{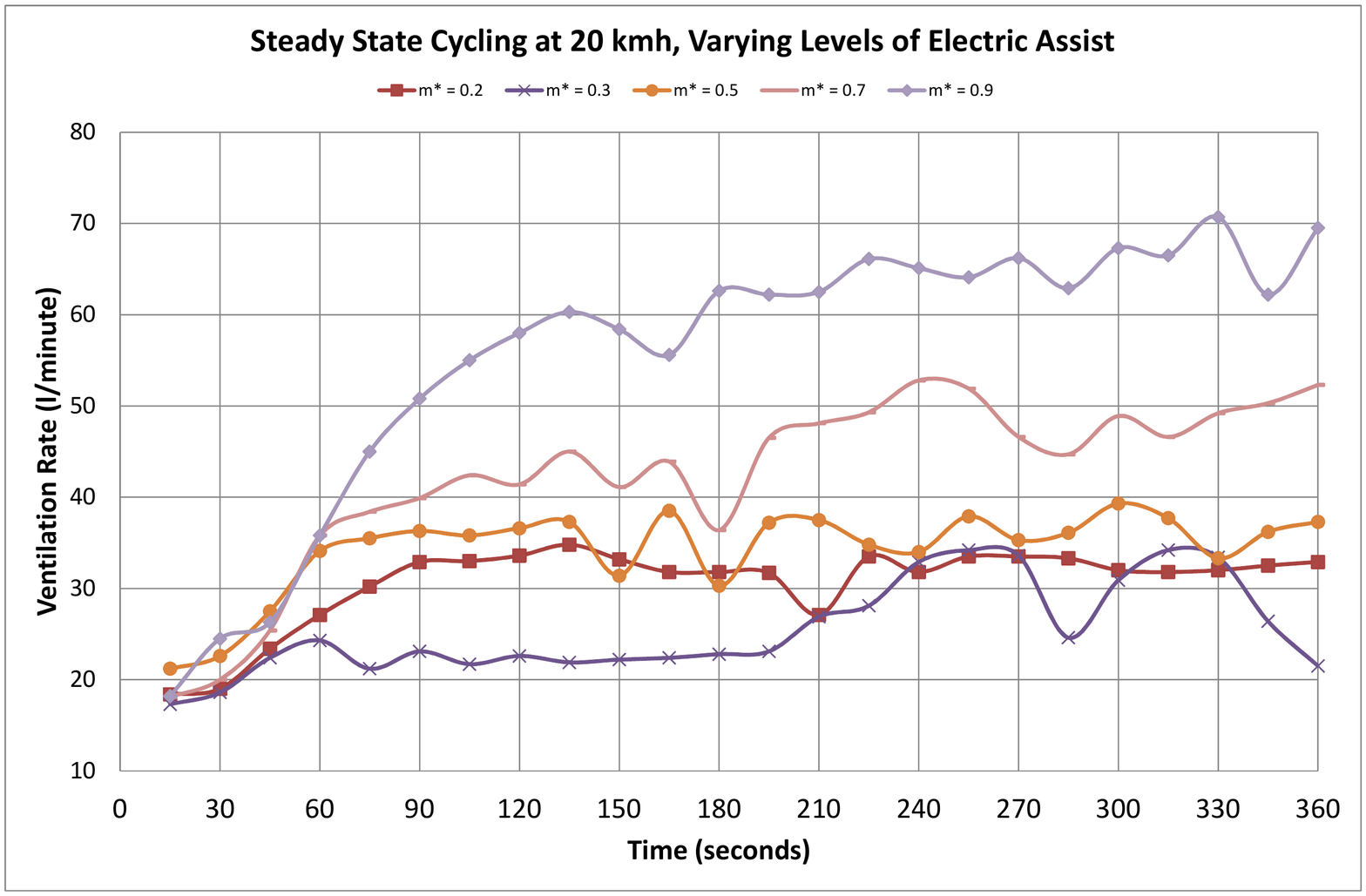}}
		\caption{Measured ventilation rate data of the test subject.}
		\label{fig:subjectA_BR}
	\end{center}
\end{figure}

\begin{figure}[h!]
	\begin{center}
		{\includegraphics[clip, trim=2.2cm 2.2cm 2.2cm 2.2cm, width=5in]{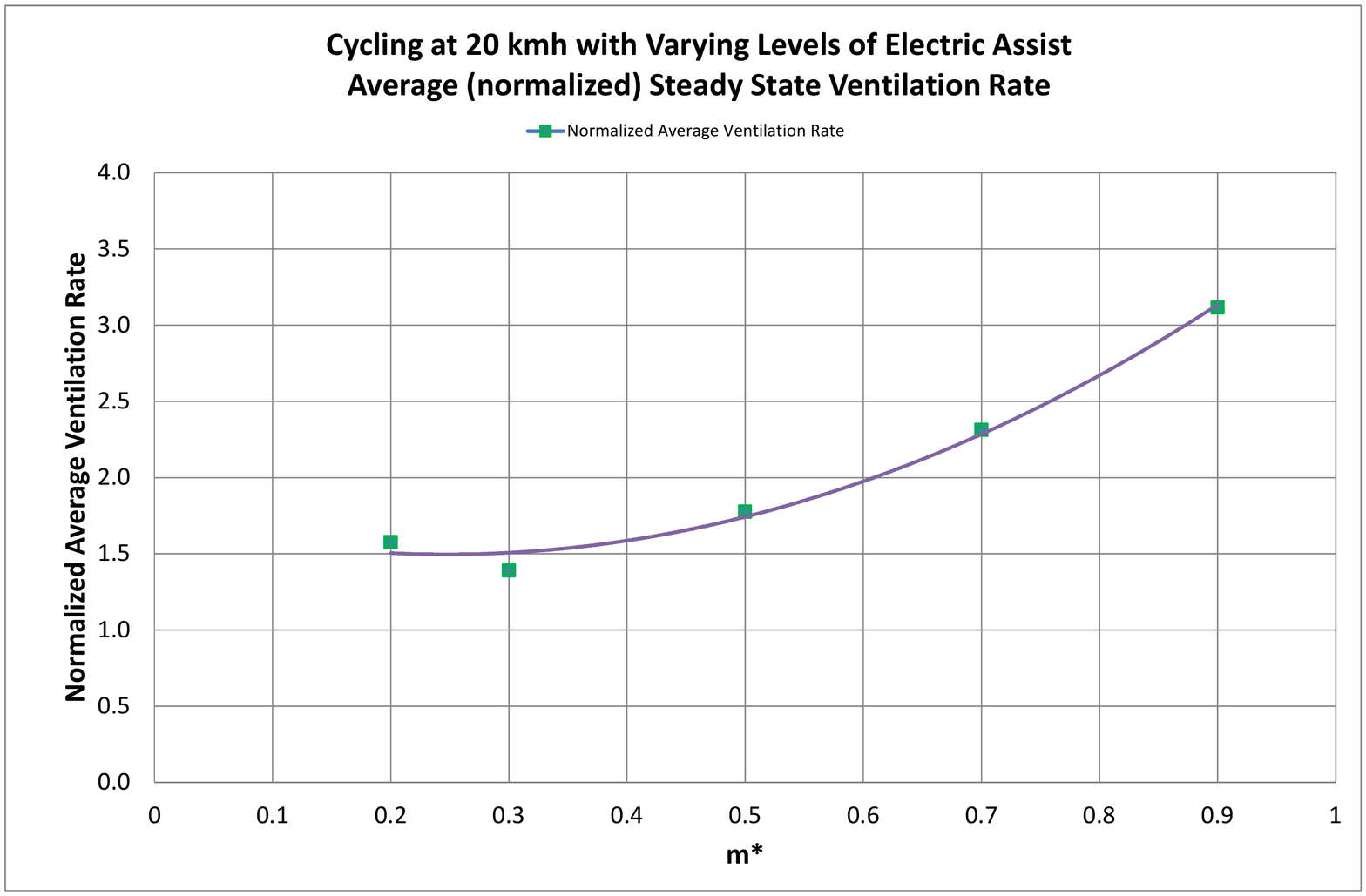}}
		\caption{Normalized ventilation rate for varying $m^*$ of the test subject.}
		\label{fig:subjectA_normalisedBRvm}
	\end{center}
\end{figure}

\begin{figure}[h!]
	\begin{center}
		{\includegraphics[clip, trim=2.2cm 2.2cm 2.2cm 2.2cm, width=5in]{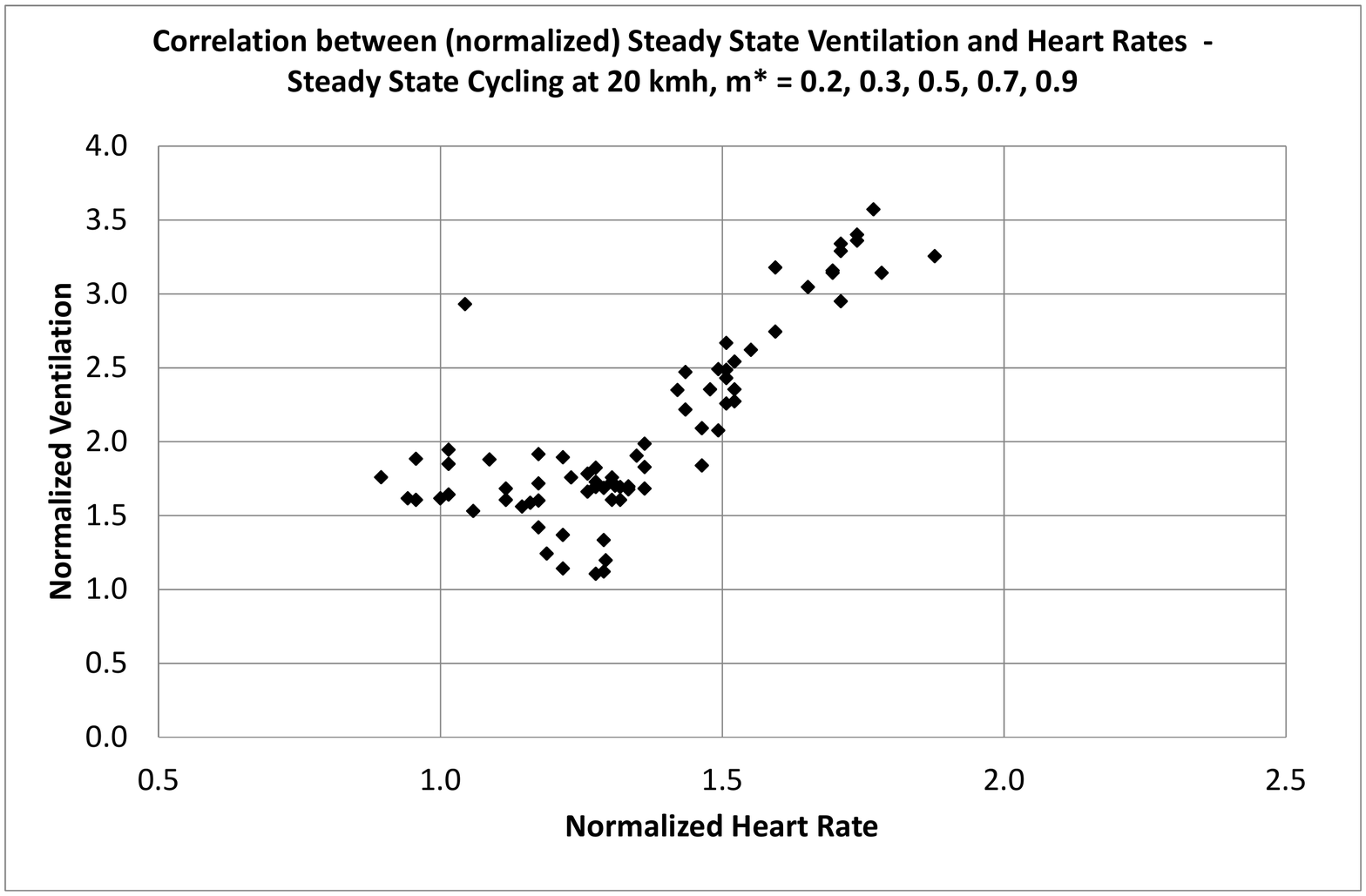}}
		\caption{Correlation scatterplot between the heart rate and the ventilation rate of the test subject.}
		\label{fig:subjectA_correlation}
	\end{center}
\end{figure}

The objective of the next test is to establish whether there is a correlation between cyclist ventilation rate and reference values $m^*$ 
across a range of fixed speeds. To this end, the cyclist was asked to maintain a fixed speed for a range of constant 
$m^*$ values during a 6 minute cycling period.  Figure \ref{fig:subjectA_BR} depicts the measured ventilation rate (starting from rest) at 20 km/h for a range of $m^*$ values. Figure \ref{fig:subjectA_normalisedBRvm} depicts these quantities normalized by resting values respectively. Finally, Figure \ref{fig:subjectA_correlation} depicts dependency of heart rate and ventilation rate.
While the instantaneous measurements show considerable variability, the average measurements clearly illustrate a relationship between $m^*$ and the ventilation rate respectively, as well as a strong correlation between heart rate and ventilation rate.  Note that this relationship  was also observed across a population of cyclists (though not depicted here).  

\subsection{Cyclist response times}

\begin{figure}[h!]
	\begin{center}
		{\includegraphics[clip, trim=2.2cm 2.2cm 2.2cm 2.2cm, width=5in]{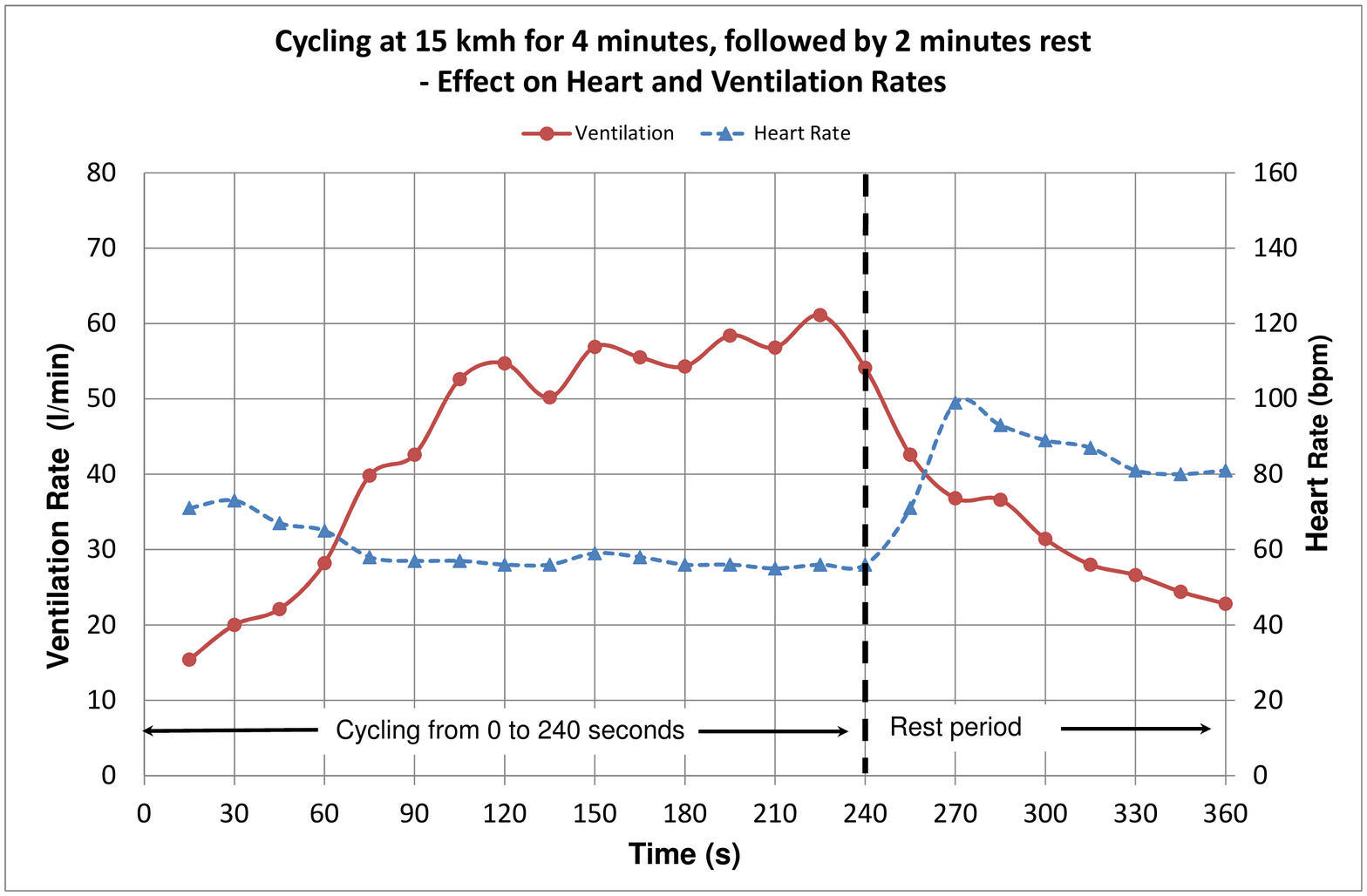}}
		\caption{Response time test at 15 km/h. The cyclist stops cycling after 4 minutes. It is observed that their ventilation rate drops from 55-60 l/min when cycling to 25-30 l/min in the 2 minutes after stopping. This is a drop of approximately 50\%.}
		\label{fig:ST_F}
	\end{center}
\end{figure}

An important consideration in the design of our cyber-physical system is the time it takes for a cyclist 
to reach metabolic steady state. To this end, the next experiment was designed to measure the time taken to reach this steady state 
when $m^*=1$. Figure \ref{fig:ST_F} depicts, with zero electrical assistance, the time taken for both the heart and the ventilation rate
of the test subject, to settle down when pedaling at a constant 15 km/h from a resting position.  
In the experiment, the cyclist maintains a constant speed for 3 mins, and then stops. Data is recorded 
for further 2 mins after stopping. Steady state conditions become apparent approximately after 2 mins.

\subsection{Step test}

\begin{figure}[h]
	\begin{center}
		{\includegraphics[clip, trim=2.2cm 2.2cm 2.2cm 2.2cm, width=3.3in]{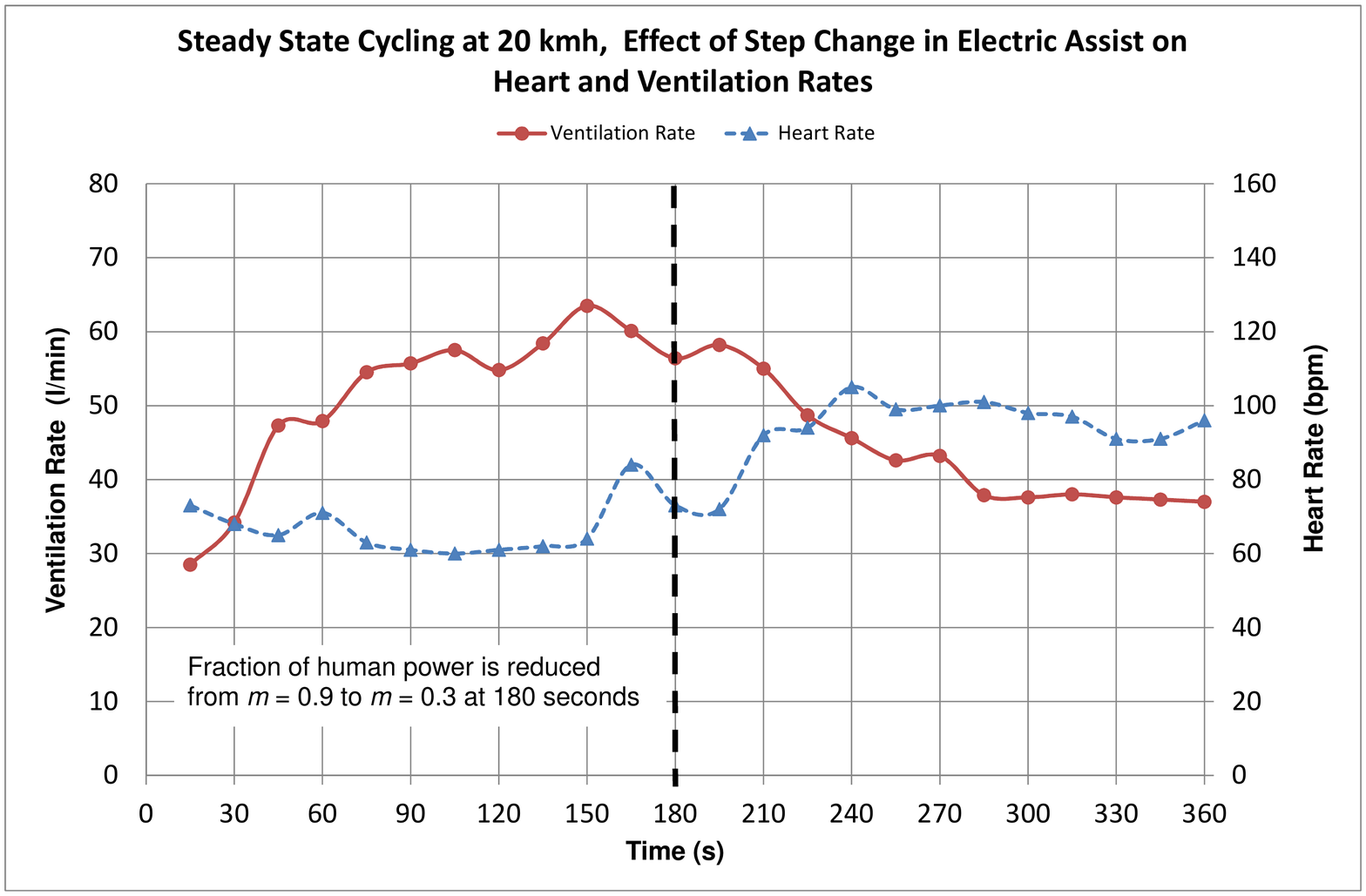}}
		\caption{Response time test at 20 km/hr. The level of electrical assistance is increased by reducing $m^*$ from 0.9 to 0.3 after 3 minutes. It is observed that the cyclist's ventilation rate reduces from approximately 55-60 l/min to 35-40 l/min. This is a decrease of approximately 35\%.}
		\label{fig:ST_C_15}
	\end{center}
\end{figure}

In this experiment, the target value of $m^*$ was varied from $m^*=0.9$ to $m^*=0.3$ at 180 seconds, while cycling at a constant speed of 20 km/h. The results of this test are shown in Figure \ref{fig:ST_C_15}.
A significant reduction in ventilation rate is observed.  Note that during transients, correlation between heart rate and ventilation rate is weak, meaning that heart rate 
is a poor proxy for ventilation rate in transient situations. This latter observation is not consistent with commonly presented medical literature, and has profound consequences 
for the use of fitness trackers and wearables in certain closed-loop applications.

\subsection{Indirect ventilation tracking}
\begin{figure}[H]
	\begin{center}
		{\includegraphics[clip, trim=0.2cm 0.2cm 0.2cm 0.2cm, width=5in]{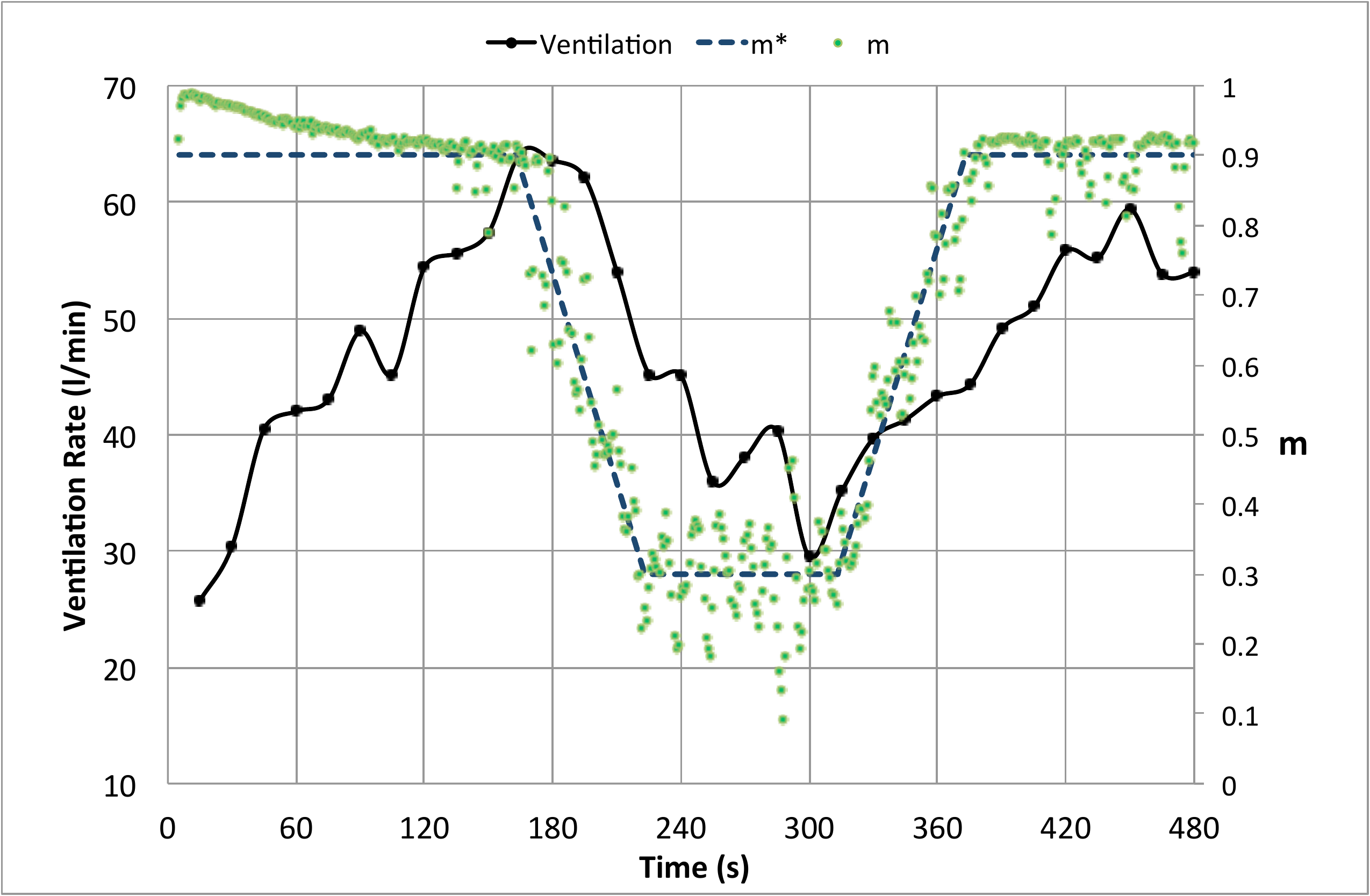}}
		\caption{Indirect control of ventilation rate.}
		\label{fig:rodrigo}
	\end{center}
\end{figure}

This experiment shows that we can indeed (indirectly) regulate ventilation rate. Figure \ref{fig:rodrigo} depicts the variation in ventilation rate as $m^*$ is varied from 0.9 to 0.3 during a 10 minute test. It can be clearly observed that the ventilation rate follows this signal smoothly with an approximate reduction of $50 \%$ from peak to minimum.

\subsection{Road Test}

\begin{figure}[H]
	\centering
	\includegraphics[width=5in]{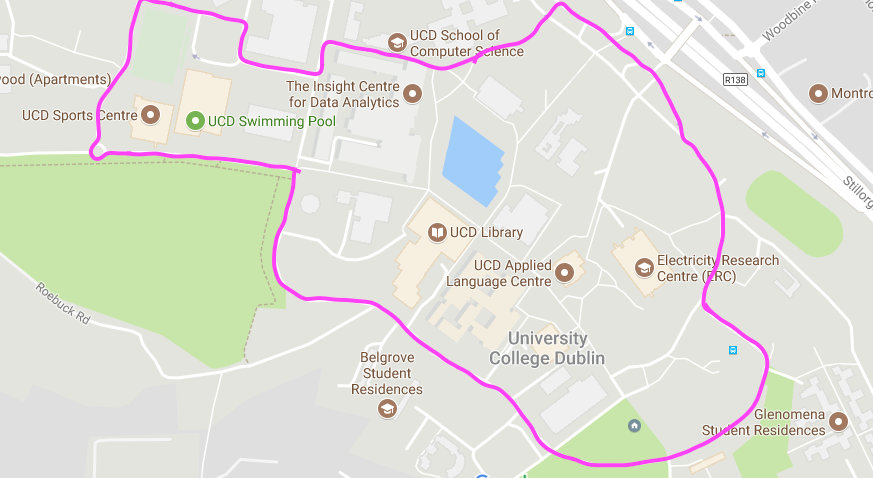}
	\caption{Route (pink line) used for road test.}
	\label{fig:ucdMap}
\end{figure}

\begin{figure}[H]
	\centering
	\includegraphics[width=5in]{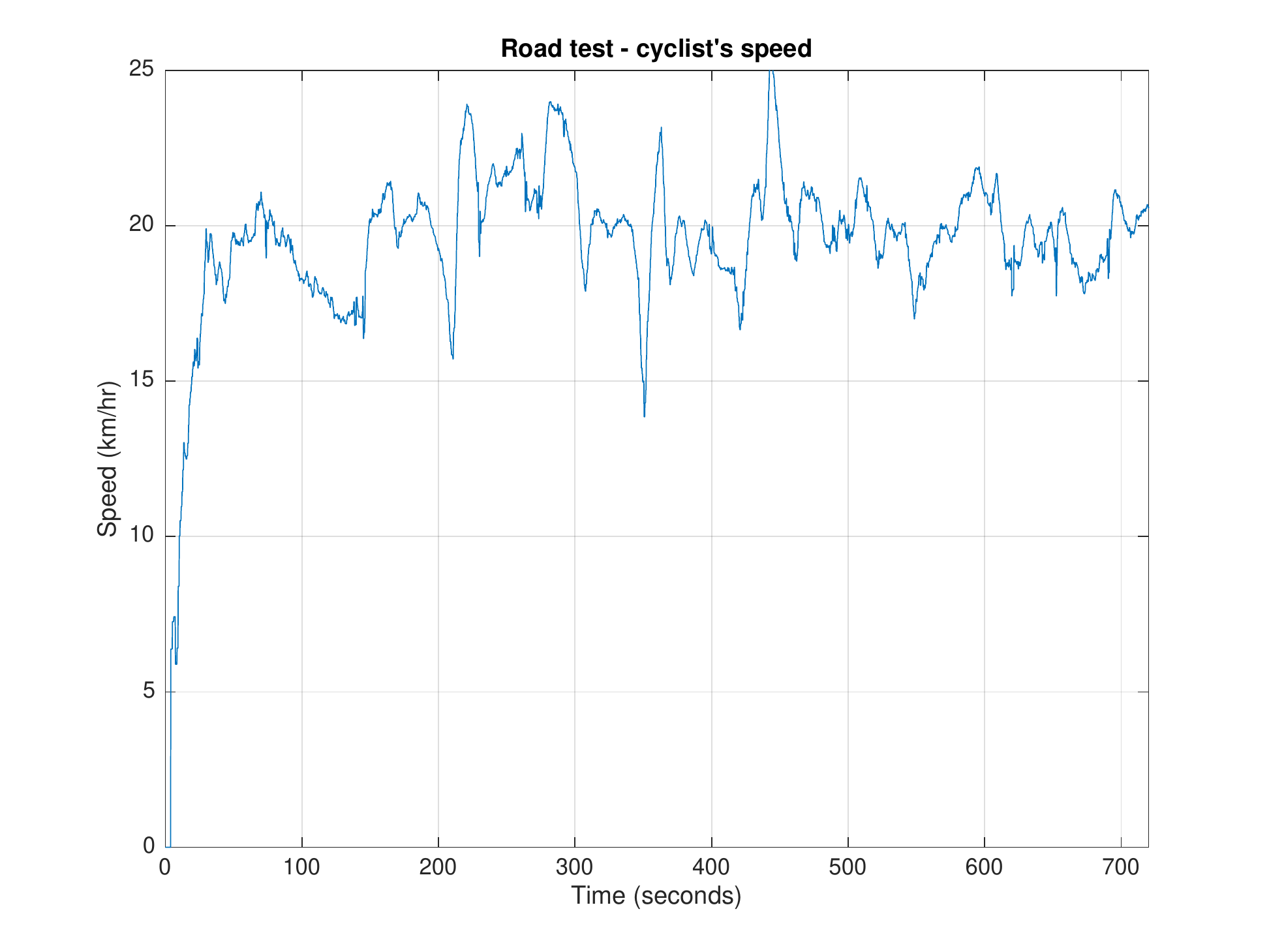}
	\caption{Cyclist's speed during road test.}
	\label{fig:cyclistSpeed}
\end{figure}

\begin{figure}[H]
	\centering
	\includegraphics[width=5in]{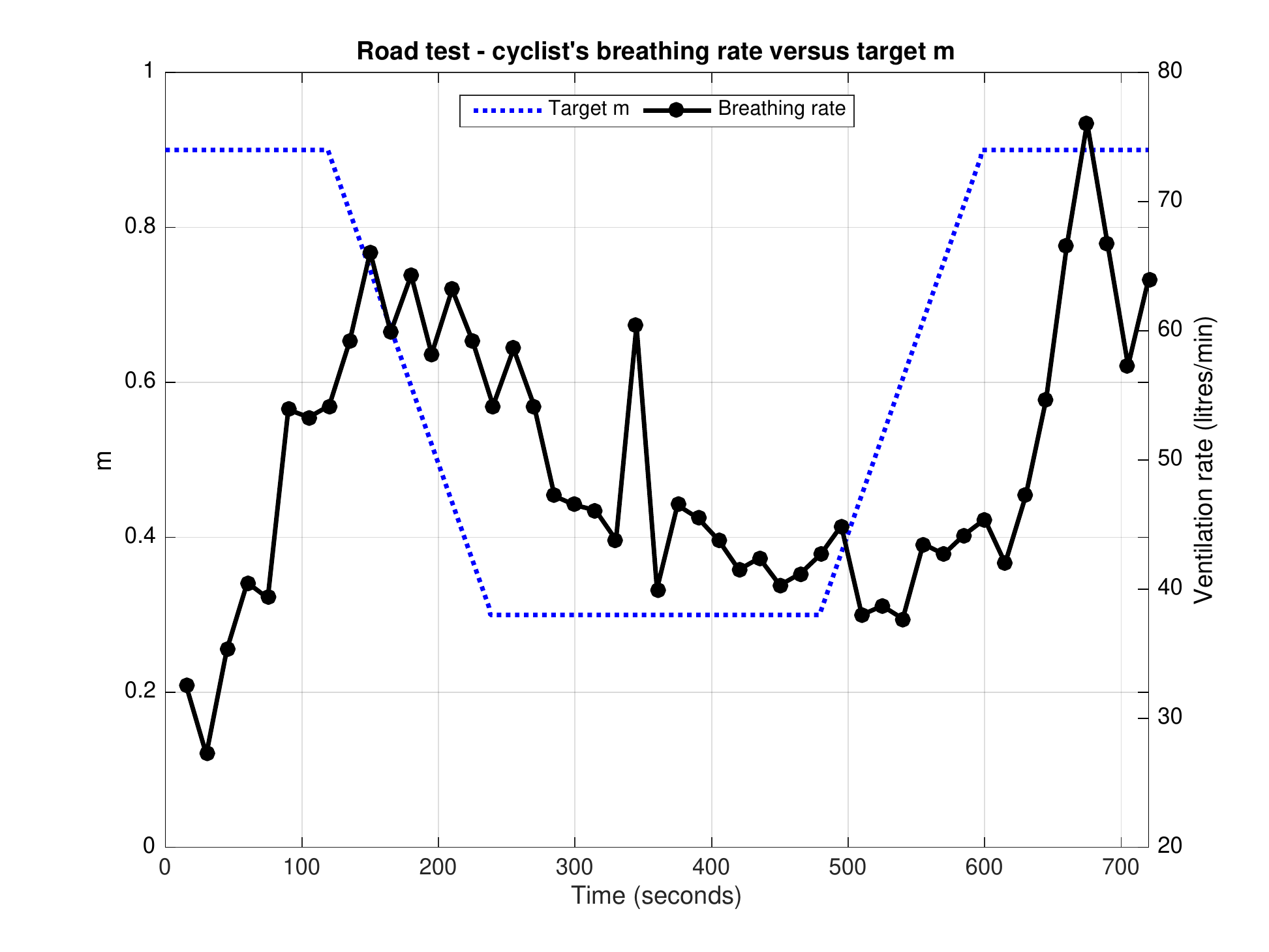}
	\caption{Control of cyclist's ventilation rate during road test.}
	\label{fig:roadTest}
\end{figure}

Our final experiment involved repeating the test while cycling around the university campus. The terrain is mostly flat. The route was 5 km long and involved completing two laps of the route highlighted in Figure \ref{fig:ucdMap}. 
The cyclist was instructed to try to maintain a constant speed of 20 km/hr. The achieved speed is shown in Figure \ref{fig:cyclistSpeed}.
As before, the cyclist's ventilation rate was monitored using spirometry equipment. Figure \ref{fig:roadTest} shows the results of the experiment. 

As was observed in the lab environment experiment, the cyclist's ventilation rate increased when the human power input was high, and decreased significantly when the human power input was replaced with power input from the electric motor. This road test experiment validated the results from the lab environment and showed that the cyclist's ventilation rate can be indirectly controlled using the control algorithms that have been presented here.

\section{Conclusions}

In this paper, we exploited the actuation possibilities afforded by e-bikes to build a new mobility service. Such a service had the overall goal of mitigating the effects of pollution on cyclists by properly providing electrical assistance. In particular, we used the actuation capabilities of e-bikes to provide electrical assistance so as to indirectly control the ventilation rate of a cyclist. Our key idea is that this indirect control can be used to reduce the air intake of cyclists in polluted areas. In order to implement our system, after having instrumented and modified an off-the-shelf electric bike, we designed a cyber-physical control system to manage the interaction of the cyclist and electric motor of the bike. A proof-of-concept implementation was validated via both lab experiments and road experiments. In both cases, the tests showed that the air ventilation rate of the cyclist can indeed be indirectly controlled via our system. While this paper shows that e-bikes can effectively be exploited to mitigate the effects of pollution on cyclists, it also opens a number of research questions that deserve to be explored in more detail. In particular, future work will include: (i) validating our approach onto a larger population to verify the scalability of our solution; (ii) further optimizing the use of the electric motor of the e-bike; and (iii) developing novel algorithms for the interaction between the cyclist and the control system.

\section* {Acknowledgments}

This work was in part supported by SFI grant 11/PI/1177. The authors are also sincerely grateful to John Gahan, Liam Carroll,
and Brian Mulkeen for assistance provided during the design phase of the project. 


\bibliographystyle{ieeetr}
\bibliography{bikes}

\end{document}